\documentclass[a4paper,fleqn,usenatbib]{mnras}

\usepackage[T1]{fontenc}
\usepackage{ae,aecompl}

\usepackage{newtxtext,newtxmath}

\usepackage{graphicx}	
\usepackage{amsmath}	
\usepackage{amssymb}	

\usepackage{float}
\usepackage{booktabs,caption}
\usepackage{threeparttable}
\newcommand{\angstrom}{\text{\normalfont\AA}}
\newcommand*\chem[1]{\ensuremath{\mathrm{#1}}}
\newcommand{\kms}{\,km\,s$^{-1}$}

\newcommand{\au}{\text{\normalfont\au}}

\title[Dwarf carbon stars are binaries]{Dwarf carbon stars are likely metal-poor
binaries and unlikely hosts to carbon planets}

\author[L. J. Whitehouse et al.]
{Lewis J. Whitehouse$^{1}$\thanks{E-mail: lewis.whitehouse.16@ucl.ac.uk},
J. Farihi$^{1}$,
P. J. Green$^{2}$,
T. G. Wilson$^{1}$,
J. P. Subasavage$^{3,4}$
\\
$^{1}$University College London, London, WC1E 6BT, UK\\
$^{2}$Harvard-Smithsonian Center for Astrophysics, 60 Garden St, Cambridge, MA 02138, USA\\
$^{3}$The Aerospace Corporation, 2310 E. El Segundo Boulevard, El Segundo, CA 90245, USA\\
$^{4}$United States Naval Observatory, 10391 W. Naval Observatory Rd., Flagstaff, AZ 86001-8521, USA
}

\date{Accepted XXX. Received YYY; in original form ZZZ}

\pubyear{2018}

\begin{document}
\label{firstpage}
\pagerange{\pageref{firstpage}--\pageref{lastpage}}
\maketitle

\begin{abstract}

Dwarf carbon stars make up the largest fraction of carbon stars in the Galaxy with $\approx 1200$ candidates known 
to date primarily from the Sloan Digital Sky Survey. They either possess primordial carbon-enhancements, or are polluted
by mass transfer from an evolved companion such that C/O is enhanced beyond unity. To directly test the binary 
hypothesis, a radial velocity monitoring survey has been carried out on 28 dwarf carbon stars, resulting in the detection 
of variations in 21 targets. Using Monte Carlo simulations, this detection fraction is found to be consistent with a 100\% 
binary population and orbital periods on the order of hundreds of days. This result supports the post-mass transfer nature 
of dwarf carbon stars, and implies they are not likely hosts to carbon planets.
\end{abstract}




\section{Introduction} \label{Intro}
Carbon stars were historically thought to lie on the asymptotic giant branch (AGB), having dredged up triple-$\alpha$ 
burning products to their surface. This gives rise to distinct atmospheric chemistry when the C/O ratio exceeds unity,
revealing strong molecular absorption bands of $\rm{{C}}_{2}$, CH, and CN. Intriguingly, it is now known that dwarf 
stars can exhibit the same distinct absorption features, thus indicating that there exists a carbon-enriched, 
main-sequence stellar population.

The first dwarf carbon (dC) star discovered was G77-61, a $T_{\textrm{eff}} \approx 4100$\,K, high proper-motion star 
that was at first assumed to be an M dwarf. A discrepancy was noticed when the $M_{\textrm{V}}=+10.08$\,mag derived 
from parallax measurements was compared to the observed colour, with the star appearing far redder than expected. 
Spectroscopy revealed strong molecular carbon features, typical of classical carbon giants, were responsible for the 
red colour \citep{Dahn} and established the first known main-sequence star with distinct C$_2$ absorption bands.

Stellar evolution does not predict the synthesis of carbon in single stars until the AGB, resulting in two possible 
explanations for the atmospheric chemistry of G77-61. The first hypothesis is that the star was formed in a 
carbon-enriched environment, and the second is that mass was transferred from an evolved companion (now unseen;
\citealt{Dahn}). Radial velocity monitoring of G77-61 over a baseline of three years revealed variations consistent with 
a circular orbit and 245\,d period \citep{Dearborn}.The mass function indicated the unseen component must have a 
mass of at least $0.55M_{\sun}$, consistent with a white dwarf.

Carbon is produced via the triple-$\alpha$ process through helium shell burning on the AGB. This material is then mixed 
into the envelope raising the carbon abundance over time via a series of convection and pulsation episodes, with the 
largest being the third dredge-up. This process can produce a C/O ratio well above unity for stars of intermediate mass 
\citep{Iben}. If the star is part of a binary, then this carbon-rich material can be transferred to the companion via Roche 
lobe overflow or efficient wind capture. However, the mass transfer mechanism is currently unconstrained for dC stars 
owing to the lack of information on orbital separations.

Mass transfer of carbon-rich material from an AGB star is widespread amongst binary systems, with other well-known 
examples being carbon-enhanced, metal-poor s-type (CEMP-s), Ba, and CH stars. CEMP-s stars are defined by their 
relatively low metallicity, high carbon abundance, and high abundance of Ba ($\rm{[Fe/H]} < -2.0$, $\rm{[C/Fe]} > +1.0$,
$\rm{[Ba/Fe]} > +1.0$; \citealt{Aoki}), whereas Ba and CH stars are more loosely defined as containing strong absorption 
features of Ba and CH respectively. These stars are typically giants with Ba and CH stars found in the red clump and on
the main-sequence turn-off respectively \citep{Esc}, while CEMP-s stars populate the first ascent giant branch (RGB). All 
three populations exhibit radial velocity variations consistent with high binary fractions and orbital periods typically on the 
order of hundreds to thousands of days \citep{McClure,Jorissen,Hansen}.

In this paper the first results of a radial velocity monitoring survey are presented for $28$ dC stars. The results to date are 
consistent with a binary fraction possibly as high as $100\%$, supporting a post-mass transfer origin. In Section \ref{target} 
the target selection and observations are described, with the results given in Section \ref{results}. The results are discussed 
in Section \ref{disc}, with the preliminary conclusions presented in Section \ref{conc}.

\section{Target selection and observations} \label{target}

Potential targets were compiled from the literature based on brightness, and selected to have a high likelihood of being 
a main-sequence star. The bulk of potential targets were found within the Sloan Digital Sky Survey (SDSS), with the first 
few hundred candidates identified via colour cuts \citep{Downes}. The largest sample of dC candidates to date was later 
discovered via cross-correlation to template spectra in DR7 and DR8 \citep{Green}. Additional dC stars identified via colour 
and proper motion were also included among potential targets \citep{Liebert,Totten,Lowrance,Rossi}, including the prototype 
G77-61. From an initial pool of 73 potential targets brighter than $g=19.0$\,AB\,mag, roughly three dozen stars have been 
observed at least once, and the 28 targets with two or more observations are listed in Table \ref{tab:RVtab}.

\begin{figure*}
\centering
\includegraphics[width=\textwidth, height=8cm]{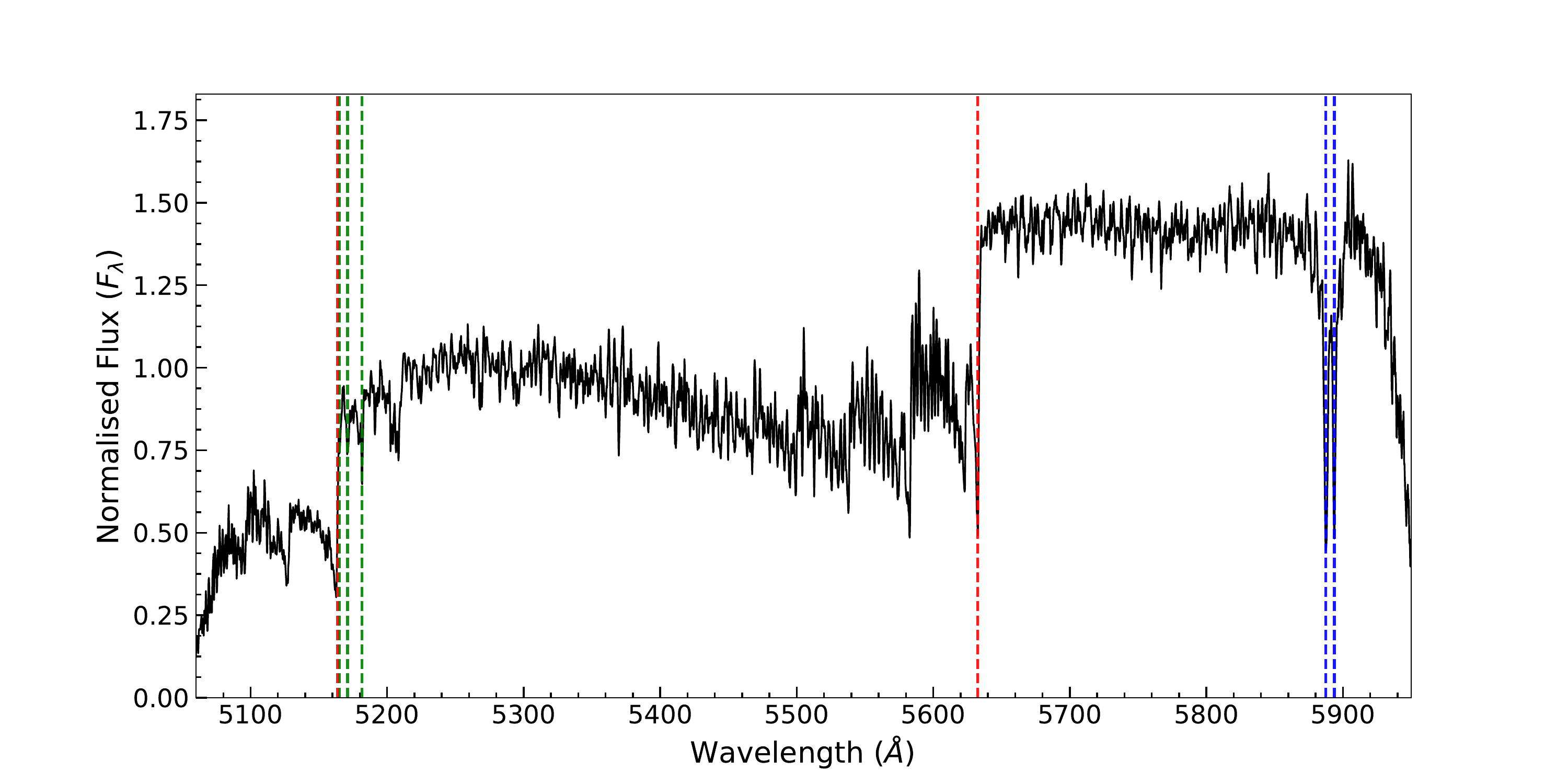}
\caption{The combined ISIS spectrum of the target LP\,225-12 showing the full spectral range for the adopted instrumental setup,
including vignetting towards the end points. The red lines correspond to the \chem{C_2} Swan band heads, green corresponds to 
the $\ion{Mg}{i}$ triplet, and blue corresponds to the $\ion{Na}{i}$ D doublet.}
\label{2m1622_spec}
\end{figure*}

The observations were carried out using the ISIS spectrograph on the William Herschel Telescope at Roque de los Muchachos. 
The data reported here were obtained between 2013 February and 2017 August, using the ISIS blue arm and the EEV12 
detector with no dichroic. The R1200B grating and a $1 \arcsec$ slit were used to achieve a resolving power $R \approx 6400$
over the range $5000$ -- $6000\,\angstrom$. This choice was motivated by the presence of several strong absorption features 
for robust cross-correlation, with the region including the $\rm{C}_{2}$ Swan bands at $5165$ and $5636\,\angstrom$, the 
$\ion{Mg}{i}$ triplet at $5183\,\angstrom$, and the $\ion{Na}{i}$ doublet at $5889\,\angstrom$. The spectral coverage achieved 
with ISIS and the above settings is plotted in Figure \ref{2m1622_spec}. Observations were taken at airmasses below 1.5, and 
using individual exposure times between 300 and 1200\,s, with a goal signal-to-noise (S/N) ratio $> 10$. Three or more exposures 
were taken for each target per observation, to increase S/N, and to minimise the effect of cosmic rays and detector artifacts.

To obtain reliable wavelength solutions, arc lamps were observed for $60$\,s immediately before and after each set of 
science exposures thus correcting any flexure in the optics between pointings. The CuNe$+$CuAr lamps were chosen for 
calibration owing to the high number of features in the designated spectral range. Individual arcs were then cross-identified 
against a master arc frame that consisted of several exposures taken at the start of the night at zenith.

\section{Reduction and Analysis} \label{results}

\subsection{Data reduction and binary fraction}

The spectral images were trimmed, bias subtracted, and flat fielded using standard routines in \textsc{iraf}\footnote{IRAF is 
distributed by the National Optical Astronomy Observatory, which is operated by the Association of Universities for Research 
in Astronomy (AURA) under a cooperative agreement with the National Science Foundation.}. Individual spectra were extracted 
using the {\sc apall} package and then combined as a weighted mean to facilitate the removal of cosmic rays and increase S/N.
The wavelength calibration for each combined spectrum was obtained by taking the average wavelength solution for the arc 
frames taken directly before and after the science exposure. Subsequently each target was flux calibrated using a suitable 
standard star with few absorption lines over the observed spectral range.

Radial velocity variations were evaluated using the package {\sc fxcor} that performs a Fourier cross-correlation between an
input spectrum and a given template \citep{Tonry}. For each target, the highest S/N spectrum was chosen as a template 
against which all other observations were cross-correlated \citep{Marsh}. The velocity residuals (in \kms) of the cross-correlation 
were added in quadrature to the velocity uncertainty from wavelength calibration, yielding a total error for each pair of spectra. 
This total error depends on the S/N of the target, with higher S/N targets possessing errors $<1$\kms and lower S/N targets 
exhibiting errors up to $6$\kms, but typically less than a few \kms. Only relative radial velocities were determined in the present 
study, as individual molecular transitions depend on gas temperature and pressure. Due to the presence of only a few atomic 
lines in the spectra, absolute radial velocities will be derived and presented in a forthcoming paper.

A weighted $\chi^2$ test was used to determine if any velocity variation could be due to the measurement errors \citep{Lucatello,
Stark}. The results of the weighted $\chi^2$ test are given in Table \ref{tab:RVtab}, where $p(\chi^2|\nu)$ is the probability that 
the observed radial velocity variations are real given the errors, and $\nu$ is the number of degrees of freedom (which here is 
the number of observations). A small $p$ value thus indicates a low probability that the null hypothesis of constant radial 
velocity can be accepted, implying the system is likely binary.

A limit of log$_{10}$($p)<-2$ (less than 1 chance in 100) was used to reject the null hypothesis that a target was not 
radial velocity variable. By this definition, $21$ of the $28$ targets are consistent with being radial velocity variable. Despite 
prior knowledge that G77-61 and PG\,0824$+$288 are binary \citep{Dearborn,Heber}, both were included in the statistics
as the sample was selected for brightness and visibility. No radial velocity variations were detected in PG\,0824$+$288, 
but this is unsurprising given the fact it was spatially resolved at $\approx 17$\,au (corresponding to an orbital period of 
$\ga 60$\,yr; \citealt{Far}). The binary detection rate is therefore $75\%$.

\subsection{Simulations} \label{simulations}

To numerically constrain the survey sensitivity, a set of Monte Carlo simulations was run for a model population of dC stars.
The simulation was run for $10$\,$000$ stars with an initial binary fraction of $60\%$, consistent with field stars brighter than 
$M_{ \rm {v}}\approx +8$\,mag within $25$\,pc \citep{Jahreiss}, while varying the orbital parameters. The number of radial 
velocity measurements for each simulated star was chosen randomly between $2$ and $8$ to be consistent with the 
sampling of the survey. Each set of modelled radial velocities were then analysed using the same method adopted for the 
empirical data to derive a $p$ value, where errors were randomly assigned to each modelled radial velocity measurement 
with magnitudes comparable to the total errors in the survey. To measure the sensitivity, the same log$_{10}$($p)<-2$ 
detection criteria was applied.

Orbital parameters for each simulated binary system were assigned randomly, with inclination angle $i$, argument of 
pericenter $\omega$, and pericenter phase $\phi$, all being assigned from uniform distributions. The mass of each 
simulated dC star was sampled from a Salpeter initial mass distribution \citep{Salpeter} with upper and lower bounds 
set at $0.8 M_{\sun}$ and $0.2M_{\sun}$ respectively, based on masses of K and M dwarfs. Each white dwarf mass was 
sampled from a normal distribution with $\langle M_{*} \rangle = 0.63 M_{\sun} \pm  0.14$ \citep{Tremblay}.

Initially, the simulated orbital period distribution was set to $\textrm{log}_{10}{T\,\textrm{(d)}} = 4.8 \pm  2.3$ (based on 
radial velocity studies of G dwarfs; \citealt{Duq}). The orbital eccentricity was assigned depending on the period, with 
$e = 0$ for $T < 1000$\,d, and longer periods taken from a normal distribution $\langle e \rangle = 0.4 \pm 0.2$. A 
maximum orbital period of $10$\,$000$\,yr ($\textrm{log}_{10}{T\,\textrm{(d)}} = 6.6$) was adopted as systems with 
excessively long periods would not be detected given the baseline of the survey. Sampling the aforementioned orbital 
period distribution, which has a $60\%$ binary fraction, resulted in a simulated detection rate of just $16\%$. Increasing 
the binary fraction to $100\%$ raised the simulated detection rate to $27\%$, still far below the observational detection 
rate of $75\%$. Thus the orbital period distribution of nearby G dwarfs appears to be highly discrepant with the dC star 
population.

The Monte Carlo simulations were rerun using shorter orbital period distributions to better replicate the observed detection 
rate. In order to increase the simulated detection rate, circular orbits were adopted as these yielded more detections than 
eccentric orbits (by $\approx 2\%$), and only populations with $100\%$ binary fractions were simulated. A distribution of
$\textrm{log}_{10}{T\,\textrm{(d)}} = 3.6 \pm 1.8$ was tried as this represents a reasonable approximation of the periods 
observed for Ba, CH, and CEMP-s stars \citep{Jorissen,Jorissen1,Hansen}. This yielded a simulated detection rate of 
$37\%$ and still falls significantly short of the detected rate.

Shorter orbital period distributions were then explored by decreasing the mean of the log-normal distribution in steps of 
$0.2$\,dex. The standard deviation of the distribution was initially set at one half the value of the mean, and then decreased 
in steps of $0.1$\,dex. This method was carried out until the simulated detection rate converged closest to that observed.
While there is no unique period distribution that best approximates the detected binary fraction, a good representative 
distribution is $\textrm{log}_{10}{T\,\textrm{(d)}} = 2.0 \pm  0.8$, and yields a simulated detection rate of $72\%$. Most 
targets show $\Delta v_{\textrm{rad,max}}$ on the order of $20$ -- $40$\kms, thus consistent with orbital periods of 
hundreds to thousands of days. Therefore, both the simulations and the observed changes in radial velocities both 
provide consistent orbital period range estimates.

If the dC star population is $100\%$ binary, then the targets that do not show sufficient radial velocity variations are 
therefore probably either long period binaries or low inclination systems that would yield small velocity semi-amplitudes. 
Because two stars (SDSS\,J112633 and SDSS\,J120024) were only observed twice each, and with a baseline of less 
than five days each, there is a wide range of orbital periods that could not be detected for these targets. The remaining 
four null-detections (SDSS\,J074257, SDSS\,J110458, SDSS\,J130744, and SDSS\,J145725) have a baseline of over 
three years, possibly suggesting inclinations that are close to pole-on. Additional data are required to determine the 
exact orbital parameters, because the bulk of targets have an insufficient number of observations.

\begin{table*}
	\centering
	\caption{Dwarf carbon stars observed during the radial velocity monitoring survey with at least
  two observations. Column six shows the $p$ value of each target expressed logarithmically
  with an arbitrary
  lower limit placed at log$_{10}$($p) = -6.0$, corresponding to a $10^{-6}$ chance of a target
  being non-binary.
  The seventh column gives the maximum radial velocity difference between
  any two sets of observations, and
  in the eighth column this is divided by the largest error in relative radial velocity, and is thus
  a measure of the statistical significance.
  }
	\label{tab:RVtab}
	\begin{tabular}{lccccrrr} 
		\hline
		Name & RA & Dec & Epochs & Baseline	& log$_{10}$($p$) & $\Delta v_{\textrm{rad,max}}$ & Significance \\ & & & &(d)& &(\kms)& ($\upsigma$)\\

    \hline
		LHS\,1075 			& 00 26 00.48 &	$-$19 18 52.0 &	6	& 1469 & <$-6.0$  & 26.0 & 5.3 \\
		SDSS\,J012028		& 01 20 28.56 &	$-$08 36 30.9 &	5	& 1469 & $-3.0$  & 22.4 & 5.1 \\
		SDSS\,J012150		& 01 21 50.42 &	+01 13 01.4 	&	5	& 1469 & <$-6.0$ & 71.2 & 15.3 \\
		SDSS\,J013007		& 01 30 07.13 &	+00 26 35.3 	&	5	& 1469 & $-3.6$ & 28.3 & 3.8 \\
		SDSS\,J022304		& 02 23 04.43 &	+00 45 01.3 	&	5	& 1652 & $-2.5$ & 20.2 & 4.8 \\
		G77-61$^{\rm a}$				& 03 32 38.08	& +01 58 00.0 	&	8	& 1651 & <$-6.0$  & 28.9 & 4.4 \\
		SDSS\,J074257$^{\rm a}$		& 07 42 57.17 &	+46 59 17.9 	&	5	& 1037 & $-1.0$ & 7.2 & 2.9 \\
		SDSS\,J081157		&	08 11 57.14 &	+14 35 33.0 	&	6	& 1059 & $-4.6$ & 23.5 & 3.9 \\
		PG\,0824+288		& 08 27 05.09 &	+28 44 02.4 	&	5	& 1059 & $-0.1$ & 8.3 & 1.1 \\
		C\,0930-00			&	09 33 24.64 &	$-$00 31 44.5 &	6	& 1061 & $-3.9$  & 22.4 & 5.5 \\
		SDSS\,J093334		& 09 33 34.14 &	+06 48 12.6 	&	4	& 677 & $-2.6$ & 16.9 & 5.0 \\
		SDSS\,J095545		& 09 55 45.84 &	+44 36 40.4 	&	4	& 1061 & $-5.3$ & 32.6 & 5.6 \\
		SDSS\,J101548		& 10 15 48.90 &	+09 46 49.7 	&	2	& 388 & $-3.3$ & 26.2 & 6.8\\
		SDSS\,J110458		& 11 04 58.97 &	+27 43 11.8 	&	3	& 1059 & $-0.0$ & 0.9 & 0.1 \\
		KA\,2						& 11 19 03.90 & $-$16 44 49.3 &	2	& 5 & <$-6.0$ & 70.5 & 20.5\\
		SDSS\,J112633		& 11 26 33.94 & +04 41 37.7 	&	2	& 5 & $-0.4$ & 11.0 & 1.4 \\
		SDSS\,J120024		& 12 00 24.09 &	+38 17 20.3 	&	2	& 1 & $-0.3$ & 4.1 & 1.1 \\
		CLS\,50					& 12 20 00.77 & +36 48 01.7 	&	4	& 763 & <$-6.0$  & 40.3 & 7.9\\
		SDSS\,J130744		&	13 07 44.53	& +60 09 03.7 	&	3	& 1651 & $-0.7$ & 9.8 & 1.8 \\
		SBS\,1310+561		& 13 12 42.51 &	+55 55 54.6 	&	6	& 1651 & <$-6.0$ & 32.3 & 7.2\\
		SDSS\,J145725		& 14 57 25.86 &	+23 41 25.4 	&	5	& 1469 & $-1.3$ & 20.1 & 4.9\\
		CBS\,311				& 15 19 05.99 &	+50 07 02.8 	&	4	& 579 & <$-6.0$ & 46.8 & 4.2 \\
		CLS\,96					& 15 52 37.35 & +29 27 59.1 	&	5	& 1469 & $-4.7$ & 11.0 & 5.9 \\
		LP\,225-12$^{\rm a}$				& 16 22 32.86 & +42 37 54.2 	&	6	& 1469 & <$-6.0$ & 30.8 & 3.9 \\
		SDSS\,J184735		& 18 47 35.67 & +40 59 44.1 	&	4	& 1469 & <$-6.0$ & 25.2 & 6.7 \\
		LSR\,J2105+2514	& 21 05 16.54 &	+25 14 48.6 	&	6	& 1469 & <$-6.0$ & 122.4 & 27.9\\
		LP\,758-43$^{\rm a}$ 			& 21 49 37.84 &	$-$11 38 28.5 &	6	& 1469 & <$-6.0$ & 29.2 & 3.3 \\
		SDSS\,J235443		& 23 54 43.13	& +36 29 07.1 	&	5	& 1030 & $-4.7$ & 34.2 & 6.7\\
		\hline
	\end{tabular}
  \begin{tablenotes}
    \item $^{\rm a}$ These four targets possess orbital periods constrained in the literature.
  \end{tablenotes}
\end{table*}

\section{Discussion} \label{disc}

\subsection{Binary fraction and orbital evolution}

The observed variation in relative radial velocities of dC stars are consistent with the hypothesis that carbon-enhanced 
material was transferred from an evolved companion. Within the observed sample, $75\%$ show clear variations and 
are consistent with a $100\%$ binary population. The simulated period distributions that best match the observational 
detection rate suggest that either Roche Lobe overflow or efficient wind capture may be responsible for the observed 
pollution in dC stars.

To date five bona fide dC stars have determined orbital periods. SDSS\,J125017 was shown to have $2.9$\,d periodic 
variability in its lightcurve, and this was confirmed to correspond to its orbital period via radial velocity measurements 
\citep{Margon}. Interestingly, in their search of $\approx 1000$ dC star lightcurves using Palomar Transient Factory 
data, \citet{Margon} find only SDSS\,J125017 exhibited variations consistent with a short period binary. This may 
indicate that the majority of dC stars do not have short orbital periods and are hence unlikely to be post-common
envelope systems. However, the sensitivity to binary-induced photometric variability has yet to be established for 
dC stars.

Three further dC binary orbital periods have recently been determined astrometrically at the U.S. Naval Observatory 
(USNO) and lie in the range $\approx400$--$4000$\,d \citep{Harris}. This period range is broadly consistent with the
distribution adopted in Section \ref{simulations} based on binary simulations that are well-matched to the detected
fraction of stars with radial velocity changes. Importantly, all three of these targets have at least five radial velocity
observations in this study, and their $\Delta v_{\textrm{rad,max}}$ values in Table \ref{tab:RVtab} are consistent
with Keplerian orbits at their determined inclinations. Thus, the USNO astrometry strengthens the argument that
that dC stars are likely to typically have orbits on the order of hundreds to thousands of days. 

Interestingly, together with G77-61 which possess a period of $245$\,d \citep{Dearborn}, these four dC stars 
appear to lie in a ``no man's land'' for low-mass, unevolved companions to white dwarfs, as theory predicts 
that any secondary should spiral in or spiral out of this region owing to the effects of mass loss during the AGB 
\citep{Willems}. Mass loss can overfill the AGB star Roche lobe and create a common envelope that causes
an initially close companion to inspiral due to friction \citep{Ivanova}. In contrast, if the initial binary separation 
is sufficiently large, then as mass is lost the orbital separation will increase to conserve angular momentum. 

These theoretical predictions are strongly confirmed among commonly occurring white dwarf-M dwarf binary 
systems, where there is a clear dearth of pairs with orbital separations in the region $\sim 1$ -- $10$\,au as
established via space-based imaging in the optical \citep{Far}. In contrast, there are myriad short-period ($\la 
10$\,d; \citealt{Reb}), post-common envelope systems, and long-period ($\ga 50$\,yr) widely separated white
dwarf-M dwarf systems detected by common-proper motion \citep{far05}. Comparing the spectral types of M 
dwarfs in post-common envelope systems to those in widely separated binaries, reveals no obvious differences 
\citep{Schreiber}, therefore suggesting that neither process is capable of efficient mass transfer. This is further
suppoted by the detection of just one dC star among a sample of 1600 white dwarf-red dwarf binaries identified
from the SDSS via template matching and identifying excess red fluxes via optical and near-infrared survey data 
\citep{Reb1}. Though dC stars would not be found via template matching to a white dwarf-red dwarf composite 
spectrum, it would be expected that binaries identified via a red excess to the white dwarf spectrum could
include dC stars. Their rare nature as companions to known white dwarfs is consistent with the fact that
only 9 of 1211 SDSS dC stars possessing composite spectra.

The results to date from this study indicate that these intermediate orbital periods may be typical for dC stars,
and thus are likely a key characteristic tied to their origin. If indeed most dC stars possess periods of hundreds 
to thousands of days, then they may be similar to those found for Ba, CH, and CEMP-s stars \citep{Pols,Izzard}.
Furthermore, the similarities between these polluted systems extends to metallicity, with all three populations
metal-deficient with respect to solar, most notably the CEMP-s stars. High-resolution spectroscopy has 
revealed G77-61 is one of the most metal-poor stars known ($\rm{[Fe/H]} = -4.0$; \citealt{Plez}), and preliminary 
kinematical results based on {\em Gaia} DR1 suggest that the dC population as a whole are old and metal-poor, 
with roughly $30\%$ -- $60\%$ halo members \citep{Far1}. Thus it appears that metal-poverty is important for 
C/O enhancement in dC stars.


\subsection{Carbon chemistry exoplanets}

There has been considerable interest in the existence of exoplanets that exhibit carbon dominated chemistry 
\citep{Madhusudhan}. The existence of such planets requires that the protoplanetary material be intrinsically 
enriched in carbon such that $\chem{C}/\chem{O} > 0.8$ \citep{Bond}. In this scenario, major planet-building 
materials could be predominantly carbide minerals, allowing for a SiC, TiC, graphite mantle with an Fe--Si--Ni 
core. Such planets would be chemically distinct from the rocky bodies found within the solar system. Although 
unrelated to the present study, it is noteworthy that the minor bodies that pollute the surfaces of white dwarf stars
exhibit Earth-like or chondritic C/O, with no evidence for carbon-dominated materials \citep{Wilson}.

The potential frequency of carbon-rich exoplanets depends on the space density of viable host stars \citep{Fort}. 
While dC stars are the most numerous carbon stars in the Galaxy, they are still far less abundant than their
oxygen-rich counterparts, with approximately 1:650\,000 dC stars relative to low-mass K and M dwarfs ($0.1{M}
_{\odot}< {M} <0.8\,{M}_{\odot}$; \citealt{Bochanski,deKool}. With drastically fewer potential hosts with $\rm{C/O} 
> 1$, the expected relative abundance of carbon-rich planets could be vanishing.


Assuming carbon-rich planets can and do form around host stars with C/O $>0.8$, the results presented here, 
that all low-mass, main-sequence stars in the phase space above C/O $ = 1.0$ are consistent with $100\%$ duplicity, 
therefore diminishes the possibility of single stars with C/O $ \ge 1.0$, and thus their ability to host planets. Thus, the 
available real estate for carbon planets may be dismal. However, one subgroup of CEMP stars appears to commonly 
possess both binary and single members (the CEMP-no stars; \citealt{Stark}) and therefore may contain primordial 
carbon-enhancement. If these stars are formed from carbon-enhanced nebulae, then presumably they are possible
sites for carbon-rich planets \citep{Loeb}, notwithstanding the potentially unfavourable planet hosting frequency of
metal-poor stars \citep{Fisch}.

\section{Conclusions} \label{conc}

This radial velocity monitoring survey shows that 21 of 28 ($75\%$) dC stars exhibit radial velocity variations consistent 
with duplicity. Using Monte Carlo simulations for a $100\%$ binary population with an orbital period distribution 
$\textrm{log}_{10}{T\,\textrm{(d)}} = 2.0 \pm  0.8$, the empirical ($75\%$) and predicted (72\%) detection rates are 
well matched. Thus the dC stars appear consistent with a 100\% binary population, supporting the post-mass transfer 
nature of these stars. When compared to white dwarf-M dwarf binaries, which exhibit a bimodal period distribution, the 
dC population appears to lie between the peaks, indicating that efficient mass transfer circumvents migration to short 
or long periods.

The high binary fraction of dC stars constrains the potential real estate for carbon-rich exoplanets, owing to the extrinsic 
nature of their high carbon abundance. As dC stars are the product of efficient mass transfer, the chemistry of the system 
during the planet formation phase would not reflect the chemistry of the dC star observed today. This may also be true 
for all main-sequence stars that exhibit C/O significantly above solar; if they exist (which is uncertain; \citealt{Fort,Teske}) 
such stars could be the result of binary mass transfer. It is clear from the dC stars that carbon enhancement in a 
main-sequence star is possible via binary evolution, and thus more subtle C/O enhancements may be more common 
(e.g.\ in FGK stars).

Continued radial velocity measurements for the stars in this study are necessary to determine actual orbits. Physical 
models of mass transfer -- for example Roche Lobe overflow or wind capture \citep{Paczy,AbatePols} -- can only be 
tested with tightly constrained binary periods. State-of-the-art mass transfer models currently face challenges in 
producing carbon-enhanced stars in general \citep{Izzard,Matrozis}, and the newly uncovered dC binary population
can provide an additional and distinct set of empirical constraints.

\section*{Acknowledgements}

The authors would like to thank H. C. Harris, B. T. G\"ansicke, I. D. Howarth, and an anonymous reviewer for feedback 
that improved the quality of the manuscript. The data obtained in this paper was done so using the William Herschel Telescope
that is operated on the island of La Palma by the Isaac Newton Group of Telescopes in the Spanish Observatorio del Roque de 
los Muchachos of the Instituto de Astrof\'isica de Canarias.






\bsp	
\label{lastpage}
\end{document}